\documentclass[twocolumn,longtable]{aastex631}

\newcommand\jmh{$\mathrm{(J-H)}$}

\usepackage{booktabs} 
\usepackage{multirow} 

\shorttitle{A rotational phase dependent J-H colour of the dwarf planet Eris}
\shortauthors{Szakáts \& Kiss}

\graphicspath{{./}{figs/}}
\begin{document}

\title{Rotational phase dependent J-H colour of the dwarf planet Eris\protect\thanks{Visiting Astronomer at the Infrared Telescope Facility, which is operated by the University of Hawaii under contract 80HQTR19D0030 with the National Aeronautics and Space Administration."}}

\author[0000-0002-1698-605X]{Róbert Szakáts}
\affiliation{Konkoly Observatory,
Research Centre for Astronomy and Earth Sciences, HUN-REN, Konkoly Thege 15-17, 1121 Budapest, Hungary}
\affiliation{CSFK, MTA Centre of Excellence, Budapest, Hungary}

\author[0000-0002-8722-6875]{Csaba Kiss}
\affiliation{Konkoly Observatory,
Research Centre for Astronomy and Earth Sciences, HUN-REN, Konkoly Thege 15-17, 1121 Budapest, Hungary}
\affiliation{CSFK, MTA Centre of Excellence, Budapest, Hungary}
\affiliation{Institute of Physics and Astronomy, ELTE Eötvös Loránd University,
Budapest, Hungary}

\begin{abstract}
The largest bodies -- or dwarf planets -- constitute a different class among Kuiper belt objects and are characterised by bright surfaces and volatile compositions remarkably different from that of smaller trans-Neptunian objects. These compositional differences are also reflected in the visible and near-infrared colours, and variegations across the surface can cause broadband colours to vary with rotational phase. Here we present near-infrared J and H-band observations of the dwarf planet (136199)~Eris obtained with the GuideDog camera of the Infrared Telescope Facility. These measurements show that -- as suspected from previous J-H measurements -- the J-H colour of Eris indeed varies with rotational phase. This suggests notable surface heterogenity in chemical composition and/or other material properties despite the otherwise quite homogeneous, high albedo surface, characterised by a very low amplitude visible range light curve. While variations in the grain size of the dominant CH$_4$ may in general be responsible for notable changes in the J-H colour, in the current observing geometry of the system it can only partially explain the observed J-H variation. 
\end{abstract}


\section{Introduction} \label{sec:intro}

Light curves of small solar system bodies in the visible and near-infrared are due to the different amount of reflected sunlight at the different rotational phases of the spinning body and can be attributed to deformed shape with constant albedo, or surface albedo variegations of a spherical body. The size of the body is a key factor in this sense: asteroids larger than an effective radius of $\sim$300\,km have very little asphericity in the main belt \citep{Vernazza2021}, and therefore their light curves should predominantly originate from albedo variegations. While in general it is more difficult to infer small body shapes in the trans-Neptunian region, the shapes of for instance Pluto and Charon (large bodies with diameters of $\sim$2400\,km and $\sim$1200\,km, respectively) are very close to spherical \citep{Nimmo2017}, in a tidally locked and hence slowly rotating system. A recent analysis of the light curves of other trans-Neptunian objects shows a similar trend for large objects \citep{K23}. Multi-colour small body light curves are rare, although they could provide information on the compositional variations across the surface. Among trans-Neptunian objets a prime example is Haumea for which the difference bewtween the V and R-band light curves indicates the presence of a 'red spot' on the surface \citep{Lacerda2008}, along with a small but significant  J-H colour variation, which is consistent with the location of the spot \citep[][]{Lacerda2009}. Some surface materials (tholins, pyroxene, etc.) can unambiguously be related to visible colours which is also the basis of the canonical asteroid taxonomy \citep{2009Icar..202..160D}. At the same time other materials, like ices on low temperature surfaces in the outer solar system, have similar reflectivities in the visible, but have absorption bands at various wavelengths / have different depths in the near-infrared, leading to a variety of NIR colours \citep{FV2021}. The observations of Pluto's surface with New Horizons show how the different surface features can be connected to the near-IR spectral variations \citep[see][for a summary]{2019rca..book..442C}. 

Recent works determined that the rotation of the dwarf planet Eris is tidally locked and synchronized with the orbital period of its satellite, Dysnomia \citep{Szakats2023,Bernstein2023}, with an orbital/rotation period of P\,$\approx$\,15.8\,d. The visible range light curve is shallow with a peak-to-peak amplitude of $\Delta m$\,=\,0.03\,mag, and the light curves observed at different visible bands are consistent with constant visible range colours through the different rotational phases. While no full light curve has been measured for Eris in the near-infrared, the existing J, H and K-band measurements which sampled the light curve at random phases show large variations, most prominently in the J-H colour \citep{Szakats2020}, ranging from \mbox{J-H\,=\,$-0.290\pm0.045$} to \mbox{J-H\,=\,0.287$\pm$0.114} (in the 2MASS system). The reason behind this near-infrared colour variation could be a compositional variegation across the surface.  
Eris' near-IR spectrum is known to be dominated by strong CH$_4$ absorption bands in the near-infrared, close to the H photometric band, as shown in \cite{AlvarezCandal2011,AlvarezCandal2020}. These VLT/X-Shooter spectra were taken at two epochs with a difference of 45.9\,days, corresponding to 2.91 rotations, i.e. looking almost at the same sub-observer longitude, and indeed showing similar spectral features. In a recent paper \citet{Grundy2023} presented near-infrared spectra of Eris obtained with the NIRSpec instruments of the James Webb Space Telescope at a somewhat different orbital phase, mainly focusing on the D/H isotopic ratios. While multi-epoch deep NIR spectra are difficult to be obtained for a target like Eris, visible-NIR colours can effectively be used to identify the main constituents that determine the spectrum \citep{FV2021}. 

In this paper we present near-infrared J and H-band observations of Eris with the Guidedog Camera of the Infrared Telescope Facility\footnote{Visiting Astronomer at the Infrared Telescope Facility, which is operated by the University of Hawaii under contract 80HQTR19D0030 with the National Aeronautics and Space Administration.}, compare them with previous J-H measurements, analyse their orbital phase dependence, compare the observed colours with those of other trans-Neptunian objects, and perform some simple calculations to try to explain the J-H variations. 

\section{Observations and data reduction}\label{sec:obs}

We used the SpeX \citep{Rayner2003} GuideDog camera between August 11, 2023 and August 25, 2023 with two day cadence (see the observation log in Table \ref{tbl:obslog}). Every second day only J band observations were planned for the target. Because the field of view of the GuideDog camera is very small (1\arcmin) we observed comparison stars separately, within a few degrees to Eris. On three out of the eight nights there were no, or only short observations due to bad weather, and on the second night (August 13) because of user error. On the first two nights we used the off-axis guider to guide the telescope, but due to the lack of bright stars we switched to MORIS\footnote{\url{http://irtfweb.ifa.hawaii.edu/~moris/}}. 

\begin{figure}[ht!]
\centerline{\includegraphics[width=\columnwidth]{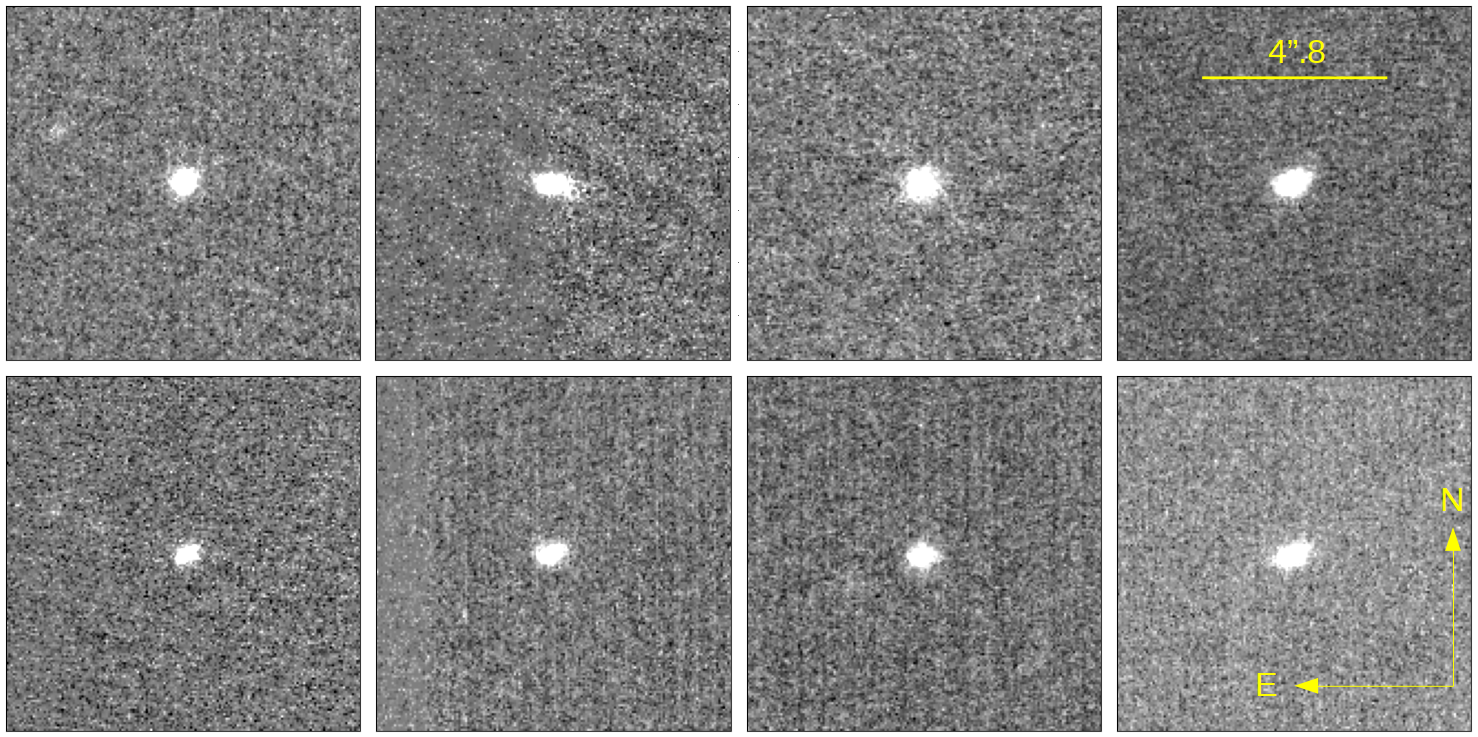}}
\caption{Cutouts of the processed frames around (136199) Eris. Top row: frames in J-band. Bottom row: frames in H-band. Every frame corresponds to a successful observing night (Table \ref{tbl:obslog}.), where both bands (J, H) was used to osbserve the target. \label{fig:frames}}
\end{figure}

In order to process the data first we made a sky image for the target and for the comparison stars (see Table \ref{tbl:obslog}.) in each filter for every night by creating the median of the five dither positions, and scaling the individual frames to a reference level. This was necessary because most of the observations were made before sunrise and the sky background got incrementally brighter with every exposure. Then we scaled and subtracted these sky images from each frame at every dither position. Finally we fitted a Gaussian center for the target on each frame and we shifted the images to a reference image (the first one in the sequence) and we took the median. The images were scaled with the exposure time, i.e. every frame which was used for photometry, was scaled to 1 second exposure time. Some cutouts of the processed images, centred on the target, are presented in Fig.~\ref{fig:frames}.

Aperture photometry was performed with Astropy and Photutils \citep{photutils2023}. We used twelve different aperture sizes ranging from 2 pixels to 13 pixels and a sky annulus of $r_{in}=23$px, $r_{out}=33$px. To estimate the background error we used six apertures with the same size of the photometric aperture around the target. Then we estimated the 1 sigma error as the standard deviation of the flux values in the background apertures.
% as seen in Fig. \ref{fig:apers}.

% \begin{center}
\begin{table*}[t]%
\caption{Observation log.\label{tab1}}
% \centering
\begin{tabular*}{500pt}{@{\extracolsep\fill}lcccr@{\extracolsep\fill}}
\toprule
\textbf{Date} & \textbf{Granted time (HST)} &\textbf{Success (Y/N)} & \textbf{Used filters} & \textbf{Comparison stars}\\
\midrule                   
\hline
\multirow{2}{*}{20230811} & 03:30-06:15 &{Y} & {J, H, (K)} & 2MASS01461142-0051514\\
& & & &2MASS01470017-0047565 \\
\hline
20230813 &05:00-06:15&  N &  -& - \\
\hline
\multirow{3}{*}{20230815} & 02:15-05:00& {Y} & {J, H, (K)} & 2MASS01475460-0044315\\
 & & & &2MASS01473713-0047366\\
 & & & &2MASS01461908-0036034\\
\hline
\multirow{3}{*}{20230817} &05:00-06:15 & {Y} & {J, (H), (K)} & 2MASS01475460-0044315\\
 & & & &2MASS01473713-0047366\\
 & & & &2MASS01461908-0036034\\
\hline
20230819 & 03:30-06:15 &N & - & - \\
\hline
20230821 & 03:45-05:00& N & - & - \\ % \cline{5-5}
\hline
\multirow{3}{*}{20230823} &03:30-06:15 & {Y} & {J, H, (K)} & 2MASS01475460-0044315\\
 & & & &2MASS01473713-0047366\\
 & & & &2MASS01461908-0036034\\
\hline
\multirow{3}{*}{20230825} & 03:45-06:15 &{Y} & {J, H, (K)} & 2MASS01475460-0044315\\
 & & & &2MASS01473713-0047366\\
 & & & &2MASS01461908-0036034\\
\bottomrule
\end{tabular*}
% \begin{flushleft}
\tablecomments{Bracketed filters used only for comparison stars.}
% \end{flushleft}
\label{tbl:obslog}
\end{table*}
% \end{center}

%\begin{figure}[ht!]
%\centerline{\includegraphics[width=\columnwidth]%{figs/eris_apertures.png}}
%\caption{The apertures (red) used to estimate the photometric error around (136199) Eris. These apertures had the same size as the aperture used for the photometry (green). In this example the radius is 10 px. \label{fig:apers}}
%\end{figure}

\section{Photometry results}\label{sec:phot}
After performing the aperture photometry we choose to use the 10-pixel-aperture for further analysis because the curve of growth has been flattened enough for that aperture. Differences between the 10-pixel-aperture magnitudes and those of the previous and the next apertures were on the $\sim$0.01\,mag level. We used the comparison stars to do standard calibration. First, we converted the $JHK_{s}$ magnitudes of these stars from the 2MASS filter system to the MKO filter system as the latter one was used during the IRTF observations. We applied the transformations from \citet{Cutri2003}. We calculated the shift between the instrumental magnitudes and the now MKO magnitudes per filter and per comparison star. We averaged those shifts and took the standard deviation as 1 sigma error for the shift. Then we applied the shift for the instrumental magnitudes of Eris and we calculated the 1 sigma error square as the sum of the square of the instrumental error and the square of the error from the shift.

In order to be able to compare our results with previous J-H values from the literature, we converted the J and H magnitudes of Eris to 2MASS magnitudes from the MKO system using the equations given in \cite{Cutri2003}. 
% Our measurements did not include K magnitudes so we got $K_{s}$ an average $K_{s}$ value from \citep{Fulchignoni2008}, \citep{DeMeo2009}, \citep{Perna2010}. These values varied, so we calculated the standard deviation and applied it as an added error to the conversion to acknowledge the scatter of the K magnitudes.
%When converted to the 2MASS system our J-H colours can be compared with the J-H colours obtained in previous works, as summarized in \citep{Szakats2023}. 
These previous J-H colours range from $-0.29\pm0.05$ \citep{Fulchignoni2008} to $0.287\pm0.114$ \citep{Snodgrass2010}. The lowest J-H colour values obtained in our measurements are compatible with the lowest J-H values obtained earlier, but even our largest colour index values remain  J-H\,$\leq$\,0.06\,mag. 
%(Note that the uncertainties in the 2MASS J-H colours are dominated by the uncertainties of the conversion relationships and the relative uncertainties of the colours in the MKO system are significantly smaller). 

\begin{figure*}[ht!]
%\centerline{\includegraphics[width=\columnwidth]{figs/J-H_period.png}}
\centerline{\includegraphics[width=\textwidth]{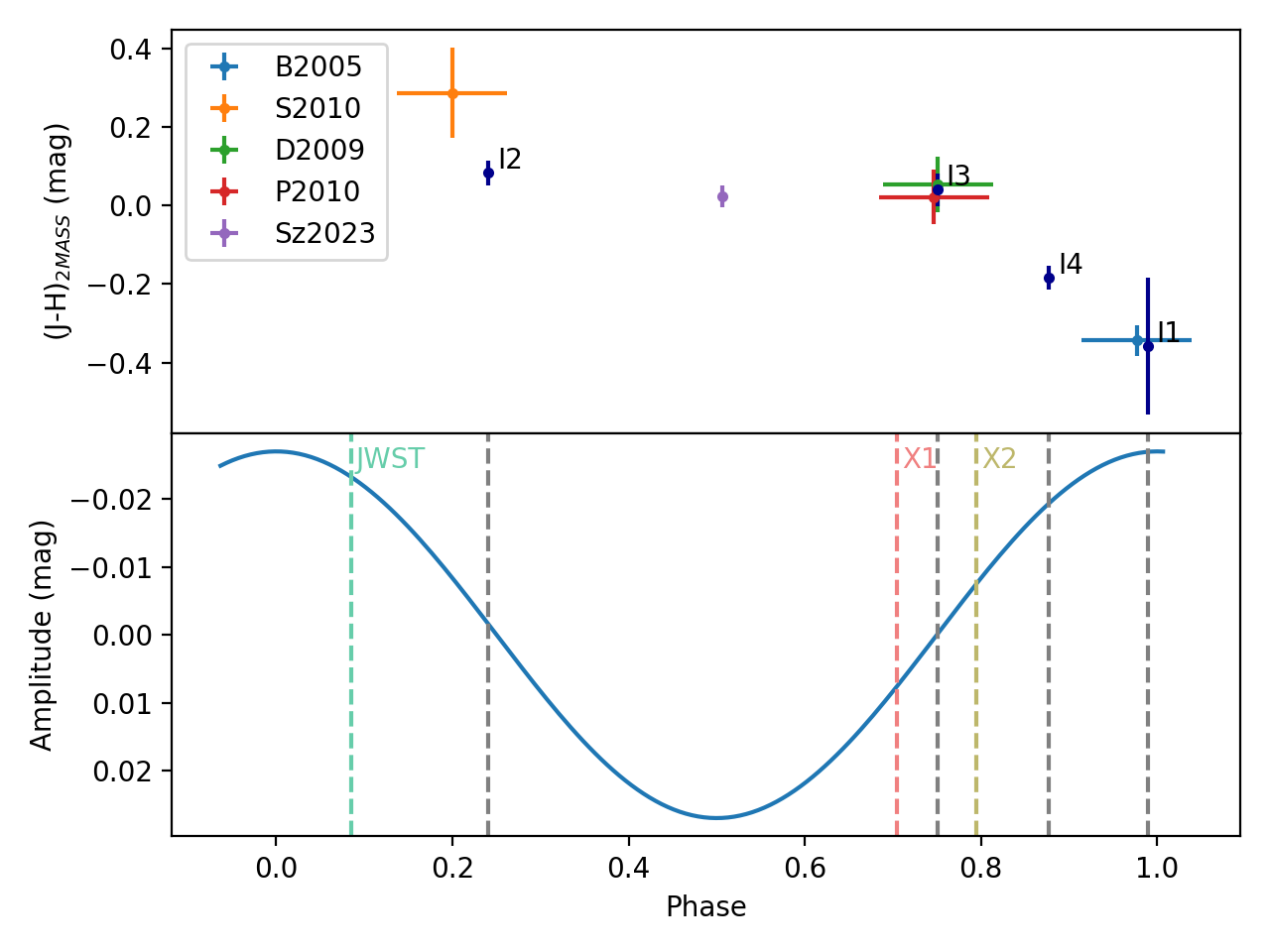}}
\caption{Top: \jmh\, colors of Eris converted to the 2MASS system, as a function of rotational phase \citep[phase zero epoch is 2457357.8929 JD, as obtained from][]{Bernstein2023}. The dark blue markers represent the J-H colors from this work, labeled as I1...I4, from Table \ref{tbl:phottable}. The colored markers represent the J-H values from the literature (Table~\ref{tbl:archive}). Bottom: An approximate representation of the Eris visible range light curve using a properly phased sinusoidal curve. The phases of IRTF, XShooter (X1 and X2) and JWST/NIRSpec (JWST) measurements are marked by vertical dashed lines. \label{fig:phase}}
\end{figure*}

% \begin{center}
\begin{table*}[t]%
\caption{Photometry results of the IRTF near-infrared measurements.\label{tab2}}
% \centering
\begin{tabular*}{500pt}{@{\extracolsep\fill}lcccr@{\extracolsep\fill}}
\toprule
\textbf{Date} & \textbf{$J_{MKO}$} &\textbf{$H_{MKO}$} & \textbf{$(J-H)_{MKO}$}& \textbf{$(J-H)_{2MASS}$}\\
\midrule                   
\hline
20230811 & 17.92 $\pm$ 0.11 &18.19 $\pm$ 0.14 &  -0.27 $\pm$  0.18 & -0.36 $\pm$ 0.18\\
\hline
20230815 & 17.96 $\pm$ 0.02& 17.91 $\pm$ 0.03 &  0.06 $\pm$ 0.04 & 0.08 $\pm$ 0.03\\
\hline
20230823 & 18.01 $\pm$ 0.03& 17.96 $\pm$ 0.02&  0.04 $\pm$ 0.04 & 0.04 $\pm$ 0.04\\
\hline
20230825 & 17.79 $\pm$ 0.02& 17.91 $\pm$ 0.03 & -0.12 $\pm$ 0.03 & -0.18 $\pm$ 0.03\\
\hline
\bottomrule
\end{tabular*}
% \begin{flushleft}
\tablecomments{Photometry results after preforming the standard calibration. We included the transformed magnitudes for the 2MASS filter system too. The transformations are based on \citet{Cutri2003}.
}
% \end{flushleft}
\label{tbl:phottable}
\end{table*}
% \end{center}

\section{Discussion and conclusions}

\begin{figure}[ht!]
\centerline{\includegraphics[width=\columnwidth]{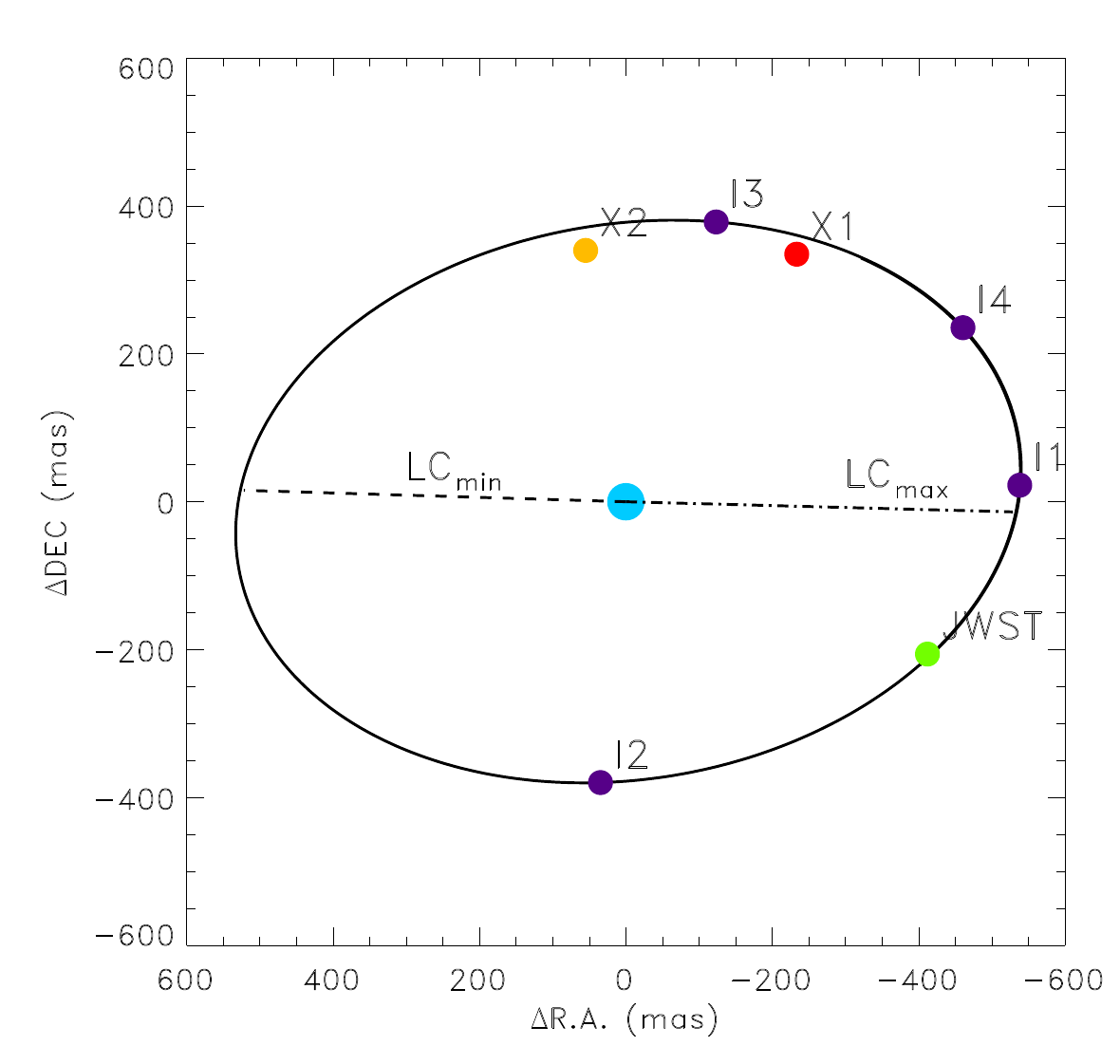}}
\caption{Orbital phasing of the IRTF measurements. 
We used the 'combined' orbit solution from \citet{Holler2021} for the epoch 2453979.0 JD: Orbital period P\,=\,15.785899$\pm$0.000050\,d; semi-major axis a\,=\,37,273$\pm$64\,km; eccentricity e\,=\,0.0062$\pm$0.0010;  inclination i\,=\,45.49$\pm$0.15\,deg; 
mean longitude at epoch $\epsilon$\,=\,125.78$\pm$0.32\,deg; 
longitude of ascending node $\Omega$\,=\,126.17$\pm$0.26\,deg; 
longitude of periapsis $\varpi$\,=\,307$\pm$12\,deg; 
The labels I1-I4 at the Dysnomia positions mark the order of the measurements in the IRTF observing sequence (see also Table~\ref{tbl:obslog}). 
Dots labeled with 'X1', 'X2' and 'JWST' mark the position and orbital phasing of Dysnomia at the time of first and second X-Shooter measurements, and the JWST/NIRSpec measurement, respectively. Dashed and dash-dotted lines mark the approximate direction of the minimum and maximum of the visible light curve, as determined by \citet{Bernstein2023}
\label{fig:orbit}}
\end{figure}

As the orbit of Dysnomia is very accurately known \citep{Holler2021} it is possible to assign an orbital phase to each IRTF observational epoch. These phases are also representative for the rotational phases of Eris due to the synchronized rotation \citep{Bernstein2023,Szakats2023}, and the actual rotational phase can also be calculated using the light curve solution presented in \cite{Bernstein2023}. 

The actual apparent relative position of Dysnomia with respect to Eris was calculated using the orbital elements of the 'combined' orbit solution in \cite{Holler2021}, as shown in Figs.~\ref{fig:phase} and \ref{fig:orbit}. 
In addition, we also calculated the orbital phases for the earlier VLT/XShooter observations \citep{AlvarezCandal2011}, and the recent James Webb Space Telescope NIRSpec measurement \citep{Grundy2023}. The XShooter and NIRSpec spectra spectra show somewhat different absorption feature depths in the wavelength range covered by the J and H photometric bands, but a direct comparison of these spectra has not been performed in \citet{Grundy2023}. 

In order to be able to compare our photometric results with those of the previous J-H, measurements (Table~\ref{tbl:archive}), we
also calculated the approximate phases using data available in the respective papers. Typically only the night of the observation is given, and in some cases the data of two consecutive nights have been merged. While this limits the accuracy of phase determination, an approximate phase even with an error of $\pm$1\,d is sufficient due to the long rotation period of Eris. This is reflected in the large phase error bars in Fig.~\ref{fig:phase}.  
Note that the J-H colour was originally obtained as J-H$_s$ in \citet{Snodgrass2010} and has a large error bar, and shows a J-H value not seen in any other NIR colour measurement of Eris. 

While the whole orbital/rotation period of Eris is not covered homogeneously by the J-H measurements, there is a clear correlation between the J-H and the light curve phase (Fig.~\ref{fig:phase}). There are two light curve phases where multiple J-H measurements are available, at $\phi$\,$\approx$\,0.75, and $\phi$\,$\approx$\,1, with three and two measurements at these phases, respectively. In the respective phases all J-H measurements consistently show the same J-H value. At a phase of $\phi$\,$\approx$\,0.2 both our I2 measurement and that by \citet{Snodgrass2010} show a similarly high J-H, however, the latter measurement also has a considerable uncertainty. J-H colors show higher values, J-H\,$\approx$\,0, around the light curve minimum, roughly between phases $\phi$\,$\approx$\,0.2-0.8, and J-H decreases towards the light curve maximum.  
%%%
% \begin{center}

\begin{table}[ht!]%
\centering
\begin{tabular}{cccc}
\hline
Epoch & (J-H)$_{2MASS}$ & (V-R) &{Reference} \\       
\hline
2453396.0 & -0.344$\pm$0.039 & 0.45$\pm$0.02 & B2005\\
%\hline
2454718.5 & 0.287$\pm$0.114	& 0.45$\pm$0.03 & S2010 \\ %\citep{Snodgrass2010}\\
%\hline
2454362.5 & 0.054$\pm$0.070	& ---  & D2009 \\ %\citep{DeMeo2009}\\
%\hline
2454441.5 & 0.022$\pm$0.070& ---    & P2010 \\ %\citep{Perna2010}\\
%\hline
2455439.8 & 0.024$\pm$0.028	& 0.358$\pm$0.030  & S2023 \\ %\citep{Szakats2023}\\
\hline
%\bottomrule
\end{tabular}
\caption{Previously obtained J-H colours of Eris, converted to the 2MASS system. These data are also used for Fig.~\ref{fig:phase}. The columns are: approximate epoch of the observations (see the text for a discussion); J-H colour in the 2MASS system; References: B2005: \citet{Brown2005}; S2010: \citet{Snodgrass2010}; D2009: \citet{Perna2010}; S2023: \citet{Szakats2023}. \label{tbl:archive}}
\end{table}

\begin{figure}[ht!]
\centerline{\includegraphics[width=\columnwidth]{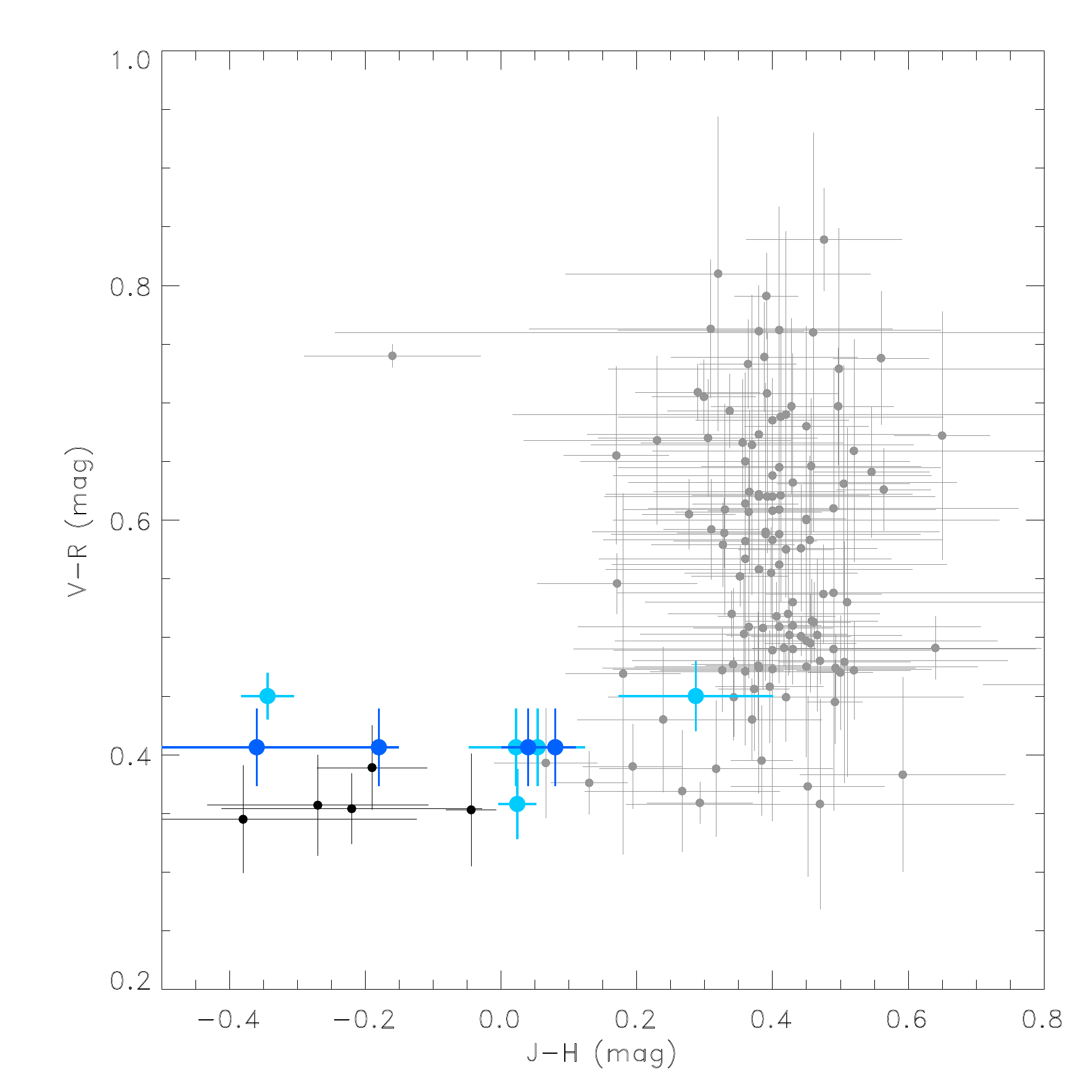}}
\caption{(V-R) vs. (J-H) colours of trans-Neptunian objects, including Eris. Gray dots with error bars represent objects from the MBOSS database, black dots are Haumea-family members. Light blue symbols correspond to the Eris colour measurements from the literature (see Table~\ref{tbl:archive}), and dark blue symbols represent the IRTF Eris measurements, assuming an average (V-R) color of 0.406$\pm$0.033. 
\label{fig:colours}}
\end{figure}

We compare the J-H colours of Eris with other trans-Neptunian objects in Fig.~\ref{fig:colours}. V-R and J-H colours were taken from the MBOSS2 database \citep{Hainaut2012}. We included Eris measurements from the literature, as well as our new IRTF measurements. We used V-R colours from the literature when it was available (see Table~\ref{tbl:archive}). However, this was not the case for all J-H measurements, and the IRTF measurements also lack corresponding V-R colours measured at the same epochs. In these cases we used a mean V-R colour of 0.406$\pm$0.033\,mag obtained from the V-R data collected in \citet{Szakats2023}. While the V-R colours of Eris are compatible with the lowest V-R values among trans-Neptunian objects, its J-H colours, observed at any rotational phase, are clearly peculiar, as very few objects show colours J-H\,$\leq$\,0.1. Most of these objects are members of the Haumea collisional family \citep[see e.g.][]{Snodgrass2010,Vilenius2018}, characterised by bright and nearly neutrally coloured surfaces (black symbols in Fig.~\ref{fig:colours}). Interestingly, the J-H colour range of Eris, J-H\,$\approx$\,--0.4 -- +0.2 is approximately the same that is covered by a number of Haumea family members (1995\,SM$_{55}$, 1999\,OY$_3$, 2005\,RS$_{43}$, 2003\,UZ$_{117}$ and Haumea in Fig.~\ref{fig:colours}).  
However, in the case of Haumea family members the colours are assumed to be due to the presence of water ice which was observed in the spectrum of Haumea \citep{PA2009}. 

\begin{figure}[ht!]
\centerline{\includegraphics[width=\columnwidth]{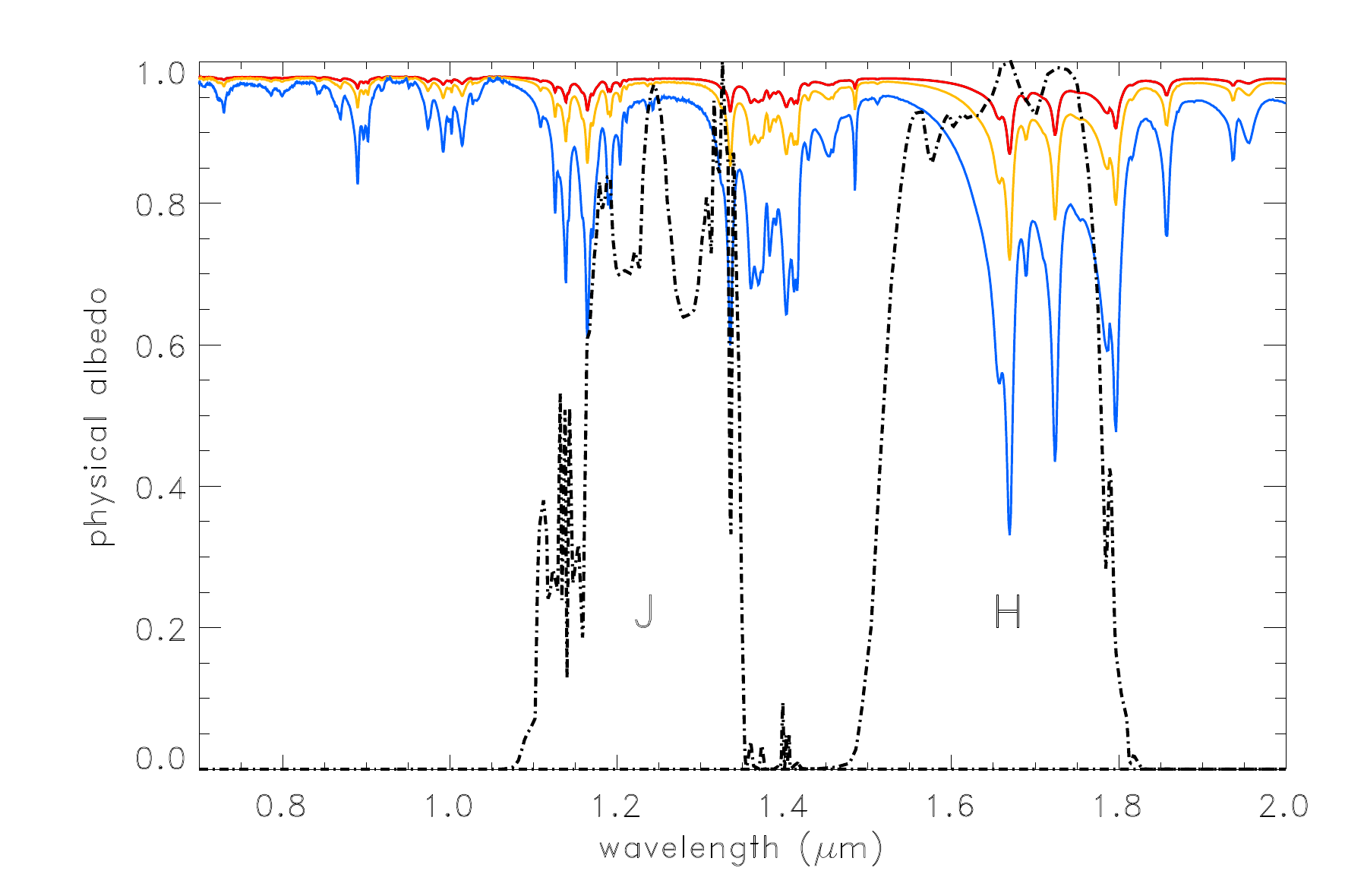}}
\centerline{\includegraphics[width=\columnwidth]{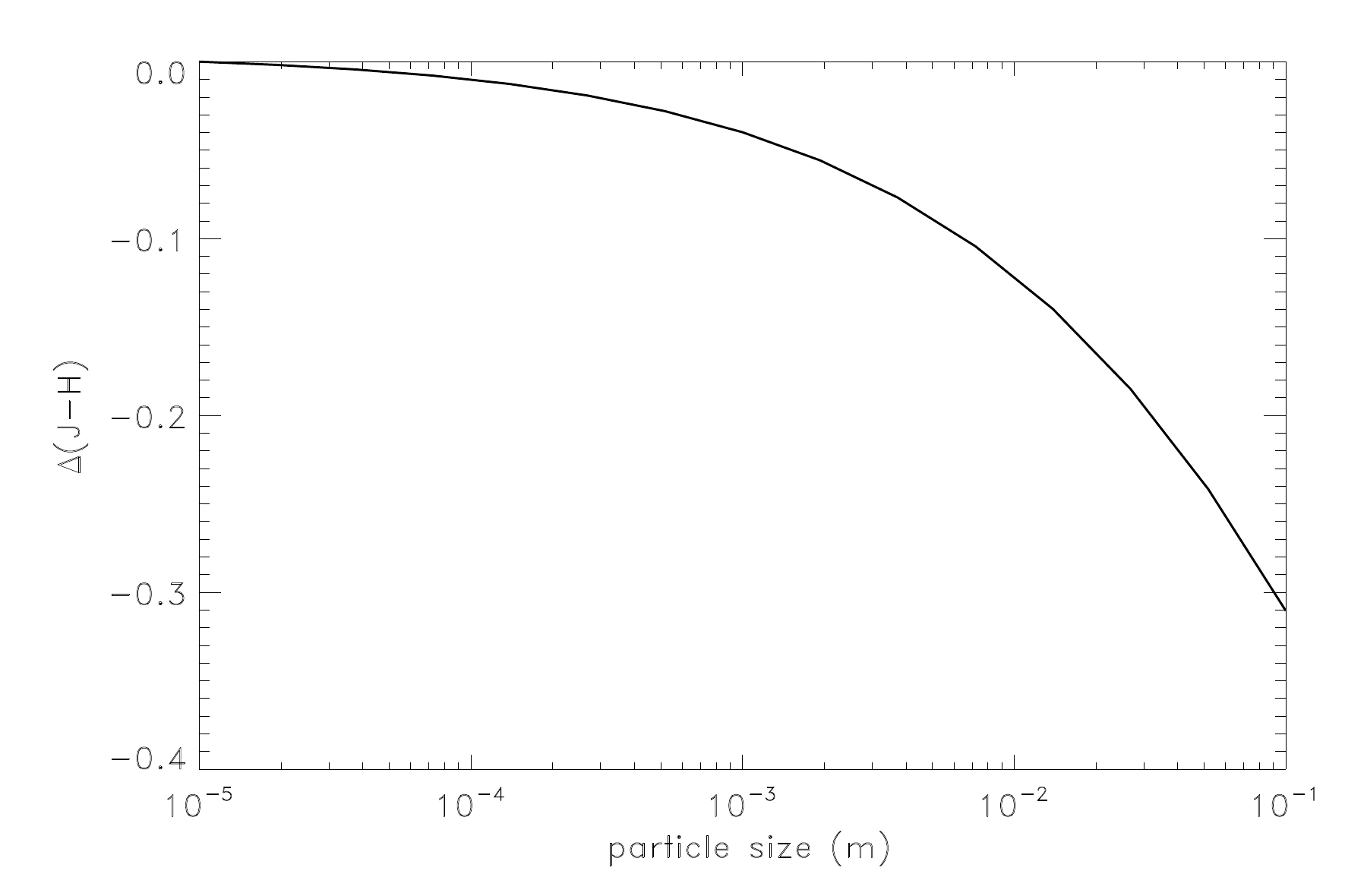}}
\caption{Top panel: Demonstration of the variation of the CH$_4$ spectrum with particle size in the 0.7--2\,$\mu$m wavelength range. Red, orange and blue curves correspond to grain sizes of 100\,$\mu$m, 1\,mm and 1\,cm, respectively. The dash-dotted curve represent the transmission curves of the 2MASS J and H-band filters.  
Bottom panel: Variation of the (J-H) colour as a function of particle size, assuming pure CH$_4$ ice, 30\,K temperature, and grains of uniform size. 
\label{fig:jmh}}
\end{figure}

In order to try to explain the variation of the J-H colour of Eris we model the visible -- near-infrared spectrum using the version of the \citet{1993tres.book.....H} scattering model presented in \citet{Trujillo2005}, applying the same parametrization. As the surface of Eris is strongly dominated by CH$_4$ ice \citep{AlvarezCandal2011,Grundy2023} we used pure CH$_4$ in the modeling. Optical constants of CH$_4$ were obtained from \citet{Grundy2002} and we used three temperatures, 20, 30 and 40\,K, and the respective absorption coefficients. We created a set of spectra using particles sizes in the range 10\,$\mu$m -- 10\,cm \citep{Quirico1999,Trujillo2005}. As demonstrated by our results (Fig.~\ref{fig:jmh}), and also in \citet{Trujillo2005} the depth of the absorption bands depends strongly on the particle size, and it is deeper for larger grains. CH$_4$ has strong absorption bands between 1.6 and 1.8\,$\mu$m, strongly affecting the H-band albedo, while the dominant part of the J-band transmittance is between two stronger absorption features and is less affected by the deepening absorption bands with increasing grain size. At the same time, the visible range albedo (in our case at $\sim$0.7\,$\mu$m) shows only a very small decrease for larger grain sizes. 
Using the J and H-band 2MASS transmittance curves \citep[see][and also in Fig.~\ref{fig:jmh}]{2mass} we calculated the expected variation of the J-H color for different grain sizes and temperatures, with respect to the J-H colour at 10\,$\mu$m grain size. The results are presented in Fig.~\ref{fig:jmh} and show a substantial decrease in the J-H colour with increasing grain size, reaching a decrease of \mbox{$\Delta$(J-H)\,=\,-0.31\,mag} at 10\,cm grain size. We present here the 30\,K curve, but the 20 and 40\,K curves are almost identical. If J-H colour variations are to be explained by grain size variations, a \mbox{$|\Delta$(J-H)$|$\,$\approx$\,0.3\,mag} change would be roughly compatible with our $|\Delta$(J-H)$|$\,$\approx$\,0.4\,mag variation found in this work, also considering the large errors on the J-H colour derived by \citet{Snodgrass2010}, as discussed above. These grain size variations would also leave visible range albedo and brightness nearly unchanged, which is compatible with the very small light curve amplitude of Eris \citep{Szakats2023,Bernstein2023}. 

A similar range of J-H colour variations can be obtained by assuming large grain sizes all over Eris' surface and adding different amount of surface constituents which have high visible range albedos but lack absorption bands in the 1-2\,$\mu$m wavelength range. The primary candidate would be N$_2$ which has just recently been detected directly for the first time on the surface of Eris \citep{Grundy2023} via its broad absorption between 4.0 and 4.3\,$\mu$m. The presence of N$_2$ has long been suspected as for CH$_4$ molecules dispersed in N$_2$ ice their vibrational absorption bands appear blue shifted, and the observed spectra suggested that the abundance of N$_2$ ice could be as high as 90\% \citep{Tegler2010,Tegler2012}; a recent spectral modeling by \citet{Grundy2023} obtained a N$_2$ abundance of 22$\pm$5\%.

However, for such large J-H variation that we observe, a very considerable hemispherical difference would be necessary on the surface of Eris, requiring that we see Eris' surface nearly equator-on. Due to the tidally locked rotation we expect that Eris' rotational pole is coincident with the pole of Dysnomia's orbit, and we see the system with a relatively large aspect angle (44\degr\, in 2023). This means that longitudes $|\vartheta|$\,$\geq$46\degr\, are continuously visible, and surface regions responsible for the varying J-H colours should be located at $|\vartheta|$\,$\leq$46\degr, relatively close to Eris' equator. In this configuration the continuously visible polar regions cover $\sim$1/3 of Eris' observer-facing hemisphere. This limits the possible J-H difference in this current geometry to $\sim$2/3 of the maximum hemispherical difference, and would limit the colour differences due to CH$_4$ grain size or compositional variations to $\Delta$(J-H)\,$\approx$\,0.2\,mag. 

Surface features with large albedo and/or compositional differences are well know to exist on Pluto \citep[see e.g.][]{2019rca..book..442C}. As discussed in \citet{Hofgartner2019} at its current heliocentric distance of $\sim$95\,au, close to its aphelion, both N$_2$ and CH$_4$ are in the local, ballistic atmosphere regime for a significant part of Eris' surface, and local, collisional atmosphere is expected only at the warmest parts, near the subsolar latitudes. \citet{Grundy2023} explained the observed D/H and $^{13}$CO/$^{12}$CO isotopic rations by a possible geologically recent outgassing from the interiors or with processes that cycle the surface methane inventory to keep the uppermost surfaces refreshed -- these processes may not occur homogeneously on the entire surface. 
While the current spectroscopic observations of Eris does not show major compositional differences at the different rotational phases, the observed variations in the J-H colour, presented in this paper, clearly indicates that there should be variegations in composition, grain size, or other surface properties that affect the near-infrared reflectance significantly. Additional visible/near-infrared spectra obtained at specific rotational phases could help to determine what effects can be responsible for the J-H colour variations. Presently the instrument most suitable to perform these kind of observations is the NIRSpec spectrometer of the James Webb Space Telescope as this is the only instrument that could directly confirm the presence N$_2$ on the surface of Eris.

\begin{acknowledgments}
Acknowledgements: This work was partly supported by the grant K-138962 of the National Research, Development and Innovation Office (NKFIH, Hungary). We thank the help by Péter Ábrahám in the reduction of the GuideDog data. We are indebted to the Infrared Telescope Facility staff for their help with the observations. This work made use of Astropy:\footnote{http://www.astropy.org} a community-developed core Python package and an ecosystem of tools and resources for astronomy \citep{astropy:2013, astropy:2018, astropy:2022}. This research made use of Photutils \citep{photutils2023}, an Astropy package for detection and photometry of astronomical sources. We thank our reviewer for the useful comments.
\end{acknowledgments}

\vspace{5mm}
\facility{IRTF(SpeX-GuideDog)}

\software{astropy \citep{astropy:2013, astropy:2018, astropy:2022}
        Photutils \citep{photutils2023}
          }

\bibliography{Wiley-ASNA}{}
\bibliographystyle{aasjournal}

\end{document}